# Data analysis of 2005 Regulus occultation and simulation of the 2014 occultation

**Costantino Sigismondi, ICRANet, UFRJ and Observatorio Nacional, Rio de Janeiro**
sigismondi@icra.it
**Tony George, IOTA North America**
triastro@oregontrail.net
**Thomas Flatrès, IOTA European Section**
thomas.flatres@wanadoo.fr

**Abstract**
On March 20, 2014 at 6:06 UT (2:06 New York time) Regulus, the 1.3 magnitude brighter star of Leo constellation, is going to be occulted by the asteroid 163 Erigone. The unusual event, visible to the naked eye over NYC, can allow to measure the shape of the asteroid, with reaching a space resolution below the diffraction limit of the eye, and of all instruments not based on interferometry. Ultimately the aperture of the instrument is related to the amount of scintillation affecting the light curve of the occultation, limiting the accuracy of video recorded data.
The asteroid profile scans the surface of the star at a velocity of 6 mas/s; the diameter of the star is about 1.3 mas and the detection of the stellar limb darkening signature is discussed, taking into consideration also the Fresnel fringes. New data reduction with R-OTE software of the 2005 Regulus occultation and simulations of the 2014 occultation with Fren_difl software are presented.

**The occultation of Regulus of March 20, 2014 above New York City**
The total solar eclipse of 1925 above New York area [5] and the occultation of delta Ophiuchi above France and Germany in 2010 [4] have been two occultation events followed by a relatively high number of observers.
Potentially the occultation of 1.3 mas [8] first magnitude star Regulus over New York City can be the first asteroidal occultation observed by million people and broadcasted on the internet [23].

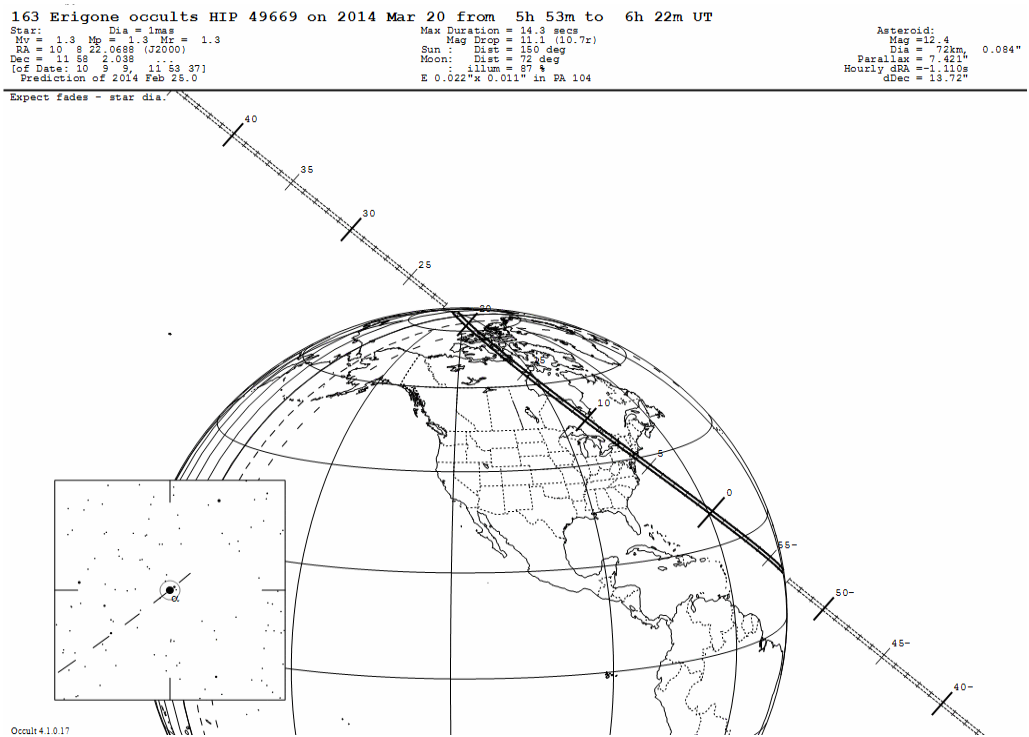

**Fig. 1** the last prediction (S. Preston [14]) on the 2014 Erigone Regulus occultation.

**The occultation of Regulus in 2005**

On October 19, 2005 in Vibo Valentia (Italy) we [1, 3] observed and recorded the occultation of Regulus with a tape Camcorder.

The Camcorder had a 3 cm objective. Later the video was digitized and posted on the web as animated gif http://www.troise.net/boliboop/occultazione-asteroidale-regulus-vs-rhodope/

and analyzed to recover the light curve of the occultation, useful to obtain the local profile of the occulting asteroid [3].

The video has been re-analyzed to using Limovie (made by Kazuhisa Miyashita http://britastro.org/asteroids/Limovie.htm ). The stellar image was surrounded by a red aperture circle, to obtain a light curve.

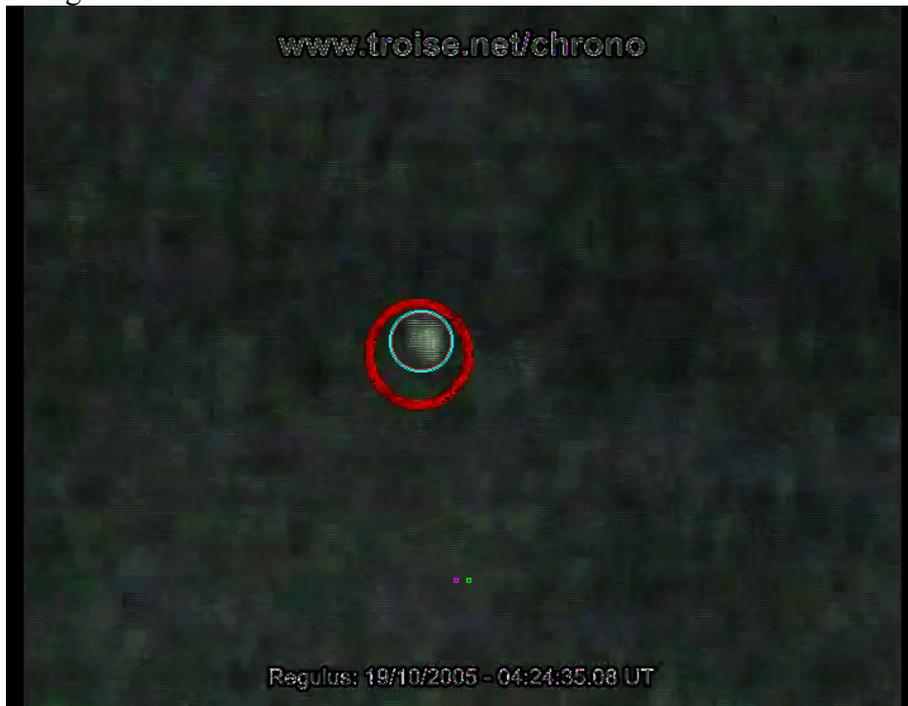

**Fig. 2** The image of the aperture used. The blue circles outline the background aperture area. The inner red circle outlines the measurement aperture area. The brightness of the video has been corrected by applying a 0.45 gamma correction. The analysis performed with Limovie used aperture photometry.

The image of the raw light curve: a cloud affected the brightness, with the occultation occurring just after the cloud started moving away from the star. This is the reason why the light curve is so much brighter on the right side. As a result, the right side of the file has been trimmed, ending up with a light curve as follows in fig. 4.

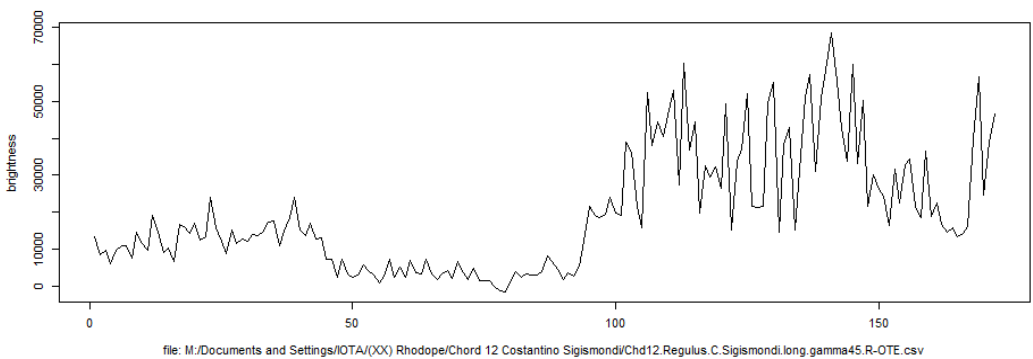

**Fig. 3** Raw light curve of Regulus occultation of 19 oct. 2005.

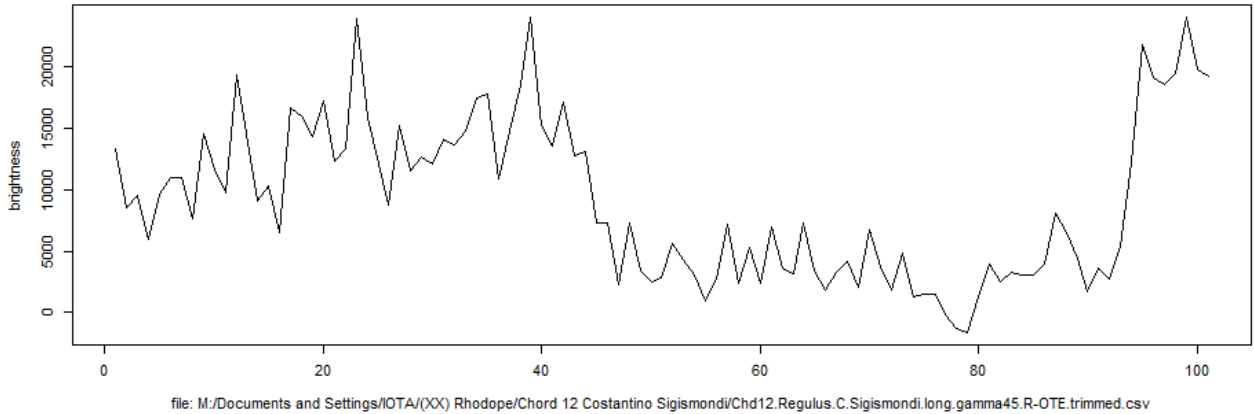

**Fig. 4** Trimmed light curve of Regulus occultation of 19 oct. 2005. The first part of the light curve has been rescaled to the last part.

Then, because there was a trend toward brightening, from the left side to the right side, a parabolic detrending curve was applied to the data, as shown below in fig. 5 and this resulted in the following trimmed and detrended light curve of fig. 6.

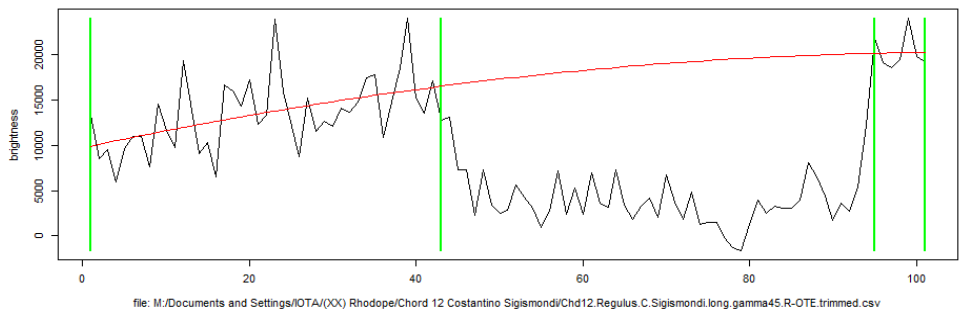

**Fig. 5** Parabolic detrend on the light curve.

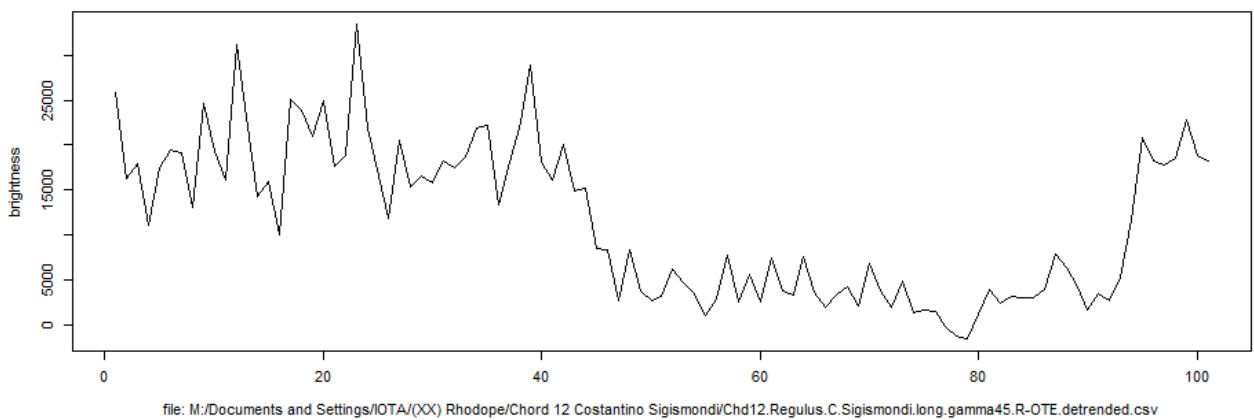

**Fig. 6** Detrended light curve. In the abscisssa are the number of the frame, at 25 frames per second.

The further analysis with R-OTE 3.1.1 software [30] is dedicated to the recovery of stellar diameter and eventually Limb Darkening signatures. R-OTE can be downloaded at http://dstats.net/download/http://www.asteroidoccultation.com/observations/R-OTE%20Release%20Package.zip

It is assumed a stellar diameter of 1.0 mas; with that stellar size, the occultation limb angles are fitted for the observed chord:
theta D (degrees) = 72.54 (+9.330/-30.820)
theta R (degrees) = 39.71 (+36.680/-39.700)
These limb angles assume no limb darkening in the stellar diameter.
Models were also run for linear law limb darkening and square root law limb darkening, but neither of these models had a better fit to this light curve than the no limb darkening model. This shows the need to reduce scintillation using a larger aperture instrument, ten times larger, for getting a signal 100 times stronger and a double frame rate, to get a signal still 50 times larger.

The duration of the event measured with this new analysis is: 1.981 (+/-) 0.044 s assuming a video rate of 25 fps.
The on-screen time stamp has been used to estimate the actual D and R times:
D disappearance (seconds) = 1.756957 (+/-) 0.030830 @ 2005-10-19 04:24:29.956957
R reappearance (seconds) = 3.738028 (+/-) 0.030830 @ 2005-10-19 04:24:31.938027
Well in agreement with the reported times. The R-OTE reported error bars of 0.04 s, much lower than the original reported error bar of +/- 1 second referred to the absolute timing
http://www.euraster.net/results/2005/index.html#1019-166 .
As this is the only video of Regulus occultation available nowadays with the second chord to receive complete reanalysis, we cannot do a full stellar diameter analysis.
The occultation of March 20$^{th}$ 2014 can permit us to get data on more chords, determining best estimates of limb angles from the best fit ellipse to the measured profile, and then processing the chords simultaneously to determine the stellar diameter of best fit to all the data.

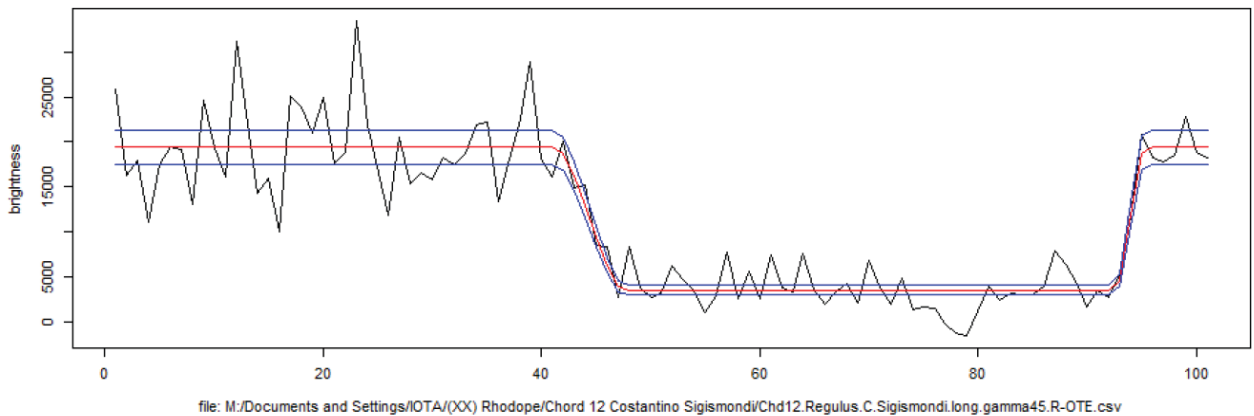

**Fig. 7** Best fit of stellar diameter on the previous light curve: 1.0 mas is the chord measured during the Vibo Valentia event. The red line is the best fit obtained with No Limb Darkening option and the blue lines represent errorbar of +/- 0.554 mas.

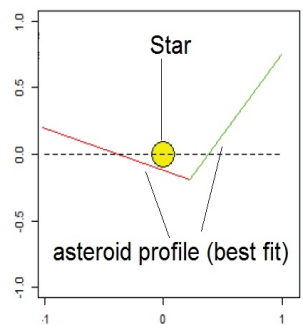

**Fig. 8** The asteroid is moving from right to left in this diagram, The red line is the disappearance edge and the green one the reappearance edge. The asteroid chord and the stellar diameter are not in scale, being 166 Rhodope about 62.5+/-3 km, about 30 mas of diameter in that occultation [1,3]. The correction to the known diameter of 166 Rhodope was a main result of the 2005 observational campaign. Here the occultation limb angles are theta(D)=73° and theta(R)=40°.

**The limb darkening scan during solar eclipses**
The spatial resolution of Limb Darkening scan with a Baily's bead light curve can attain 10 mas of accuracy for the Sun [15]. A relative accuracy of one part over 200.000 is possible for our star, but using the "Sun-as-a-star" approach [29] we can predict how the Limb Darkening signature would be observable with the whole star compressed in a milliarcsec.
The method of recovering the limb darkening from eclipse has been largely exploited in the past century [31, 32].
The duration of the phenomenon, the partial phase which is the interesting one [32] is about one hour, while the partial phase of the asteroidal occultation is of the order of a fraction of second, namely 0.2 s for Erigone-Regulus occultation.
The scintillation has been claimed as the responsible for the shadow bands [13] during an eclipse, just before the totality, while other theories have been presented [11].

**The detection of stellar companion during occultations**
Gies [19] pointed out the possibility to detect the white dwarf companion of Regulus during the occultation. The opportunity offered by the occultations, also the lunar occultations which are the fastest ones, to resolve very close stellar systems [18] is enhanced by the slow angular orbital motion of the asteroid.

**The scintillation and the electronic noise**
Modern commercial videocameras have CMOS detector, more rapid than the normal CCD videocameras. 300 fps and even 1200 fps can be operational for videos lasting 10s and more.
The extreme luminosity of Regulus allows to use these devices even without a telescope to gather enough light. The example of Venus [6] and Arcturus [7] have been presented and the incidence of the electronic noise has been discussed.
The case of Venus occultation of december 1[st] 2008 observed at 60 fps with a 21 mm objective showed a large electronic noise, which produces a scintillation-like signal, but it signature is evident when dealing with post-occultation residual signal [6,9].
With an instrument of the 20 cm class and more the influence of the scintillation is damped by a factor of $\sqrt{100}$, the electronic noise concern objects 100 times fainter. Tests with Regulus are recommended before and after the occultation.

**The limb darkening during a binary eclipse**
The method to recover the limb darkening of a star in a binary eclipsing system has been introduced by Russell and Shapley in 1912 [33, 34]. The opportunity offered by an eclipse lasting a few hours is to scan slowly the surface of the eclipsed star.
The original law of Limb Darkening, also called Russell law [35], is
*J=Jo(1-x + x•cos(i) )*
where *i* is the angle between the line of sight and the normal and x the darkening coefficient.
For the giant stars the limb darkening models have incorporated some aspect of solar physics that is not shared by giant atmospheres [36] and for these reason a square root limb darkening is hypotized:
$J=Jo(1-c_1 - c_2+ c_1 \cdot \mu +c_2\sqrt{\mu})$ with $\mu=\sqrt{1-(r/R)^2}$
Nonlinear limb darkening are required to interpret the data of beta Aurigae [10].
The "gravity darkening" of Regulus [25, 26] can be detected from the light integrated along the various chords of the stellar surface along which the asteroid profile moves.

**The limb darkening signal and the Fresnel diffraction pattern with TNO objects**
A careful study of TNO objects can be done by stellar occultations, in search of rings, dust, satellites and atmosphere [12, 21, 22].
The knowledge of Fresnel pattern [2, 20, 24] allows to model the convolution of diffraction and Limb Darkening to obtain the theoretical light curve to be compared with the observed one.
The program Fren_difl  https://docs.google.com/file/d/0BzHjNDloaOsKTi1hejJYOXN5MlU/edit authored by Thomas Flâtres can be used to this purpose.

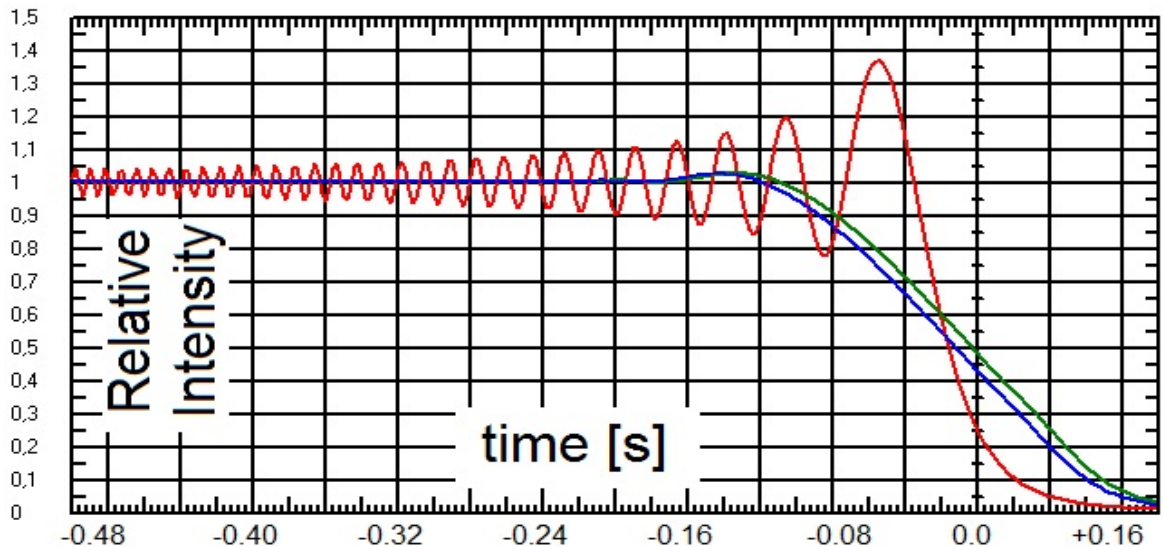

**Fig. 9** Simulation with Fren_difl of the occultation of 1.3 mas Regulus by a 84 mas asteroid 163 Erigone with 7.4" of parallax and 6 mas/s of angular velocity. The entries of the program to realize this figure are approche=24 mas/s, integration time 5 ms (200 fps), angular stellar diameter 1.300 mas. Wavelength 650 nm. The x scale of the figure has been changed of a factor of 4 to get the 6 mas/s case. The case here represented is the diametral occultation through the center of the star (with occultation limb angles theta=90°, cfr. Fig. 8); for an angle of theta=30° the approaching velocity is divided by 2 and the phenomenon is prolonged, allowing more LDF resolution on the thin partiality paths of 1745 m around the totality path, where other chords of the star are sampled.

**Conclusions**
The asteroidal occultation of one of the brightest stars in the sky allows to do detailed photometric studies on the light curve, in order to detect the signature of a faint stellar companion [19], a dust/satellite or ring distribution around the occulting asteroid [12, 22] and also the gravity darkening [25] of Regulus with a resolution comparable with the one achievable during extrasolar transits [36].
A recommended time rate of 60 fps or larger is needed to sample the diameter of the star with enough data points, typically with 6 mas/s and 1.3 mas of diameter the fading phase lasts 217 ms, with a sampling rate each 17 ms (1/60s) we have only 12 video frames.
Photometric data are much faster and more indicated for this type of study.
Telescopes of 20 cm size and more are preferred, due to the high electronic noise of commercial
 CMOS detectors like Sanyo CG9 up to 60 fps-CG10 up to 300 fps, CASIO Exilim EX-F1 up to 1200 fps, even if not produced for astronomy, they are faster than CCD and particularly adapted to this particular phenomenon.
EMCCDs Electron Multiplying CCDs http://www.andor.com/LearningPdf.aspx?id=92 in particular Photometrics Cascade 128+  http://www.photometrics.com/products/datasheets/128_.pdf up to 4149 fps, can obtain very interesting results.
Taking the risk of unconventional observations and doing tests with Regulus before the occultation, can lead the first high resolution stellar limb darkening detected during an asteroidal occultation.

Dedicated softwares to the simulation [37] and fitting [38] of occultation light curves have been here presented with their application to the occultations of Regulus in 2005 and 2014.

**Acknowledgments**
Felipe Braga Ribas, Cesare Barbieri, Andrea Richichi and Andreas Glindemann for their remarks. Arlene Werneck Barbosa for her hospitality in Copacabana, during the writing of this paper.

**References**
[1] Sigismondi, C., D. Troise and D. Montagnise, http://iota.jhuapl.edu/RegulusIOTA.pdf (2005).
[2] R. Dusser, http://www.astrosurf.com/eaon/Circulaires/Circulaire9E.htm (2005).
[3] Sigismondi, C., D. Troise, XI Marcel Grossmann Meeting on General Relativity, WSPC, 2594 (2008).
[4] http://www.euraster.net/results/2010/index.html#0708-472
[5] Brown, E.W., Astronomical Journal 37, 9-19 (1926).
[6] Sigismondi, C., arXiv:1106.2451 (2011).
[7] Sigismondi, C., J. of Occultation Astronomy 2014-1, 15 (2014) and arXiv1310.6557 (2013).
[8] Radick, R. R., Astronomical J. 86, 1685 (1981).
[9] Sigismondi, C., R. Nugent and G. Dangl, Proc. 3rd Stueckleberg Workshop on Relativistic field Theories, Pescara, Italy, 8-14 July 2008. edited by N. Carlevaro, R. Ruffini and G. V. Vereshchagin, Cambridge Scientific Publishers, p. 303 (2011).
[10] J. Southworth, H. Bruntt and D. L. Buzasi, astro-ph/0703634 (2007).
[11] Feldman, R. L., Popular Astronomy, 48, 352 (1940).
[12] Roques, F., M. Moncuquet, B. Sicardy, Astron. J. 93, 1549 (1987).
[13] Codona, J. L., Astron. Astrophysics, 164, 415 (1986).
[14] Preston, S. Sky & Tel. 127, 30 (2014).
[15] Raponi, A. et al., Solar Physics 278, 269 (2012).
[17] Rogerson, J. B., Astrophys. J. 130, 985 (1959).
[18] Sigismondi, C., Astronomia UAI 3, 30 (2005).
[19] Gies, D. R., Atel # 5917 (2014).
[20] S. Fornasier and C. Barbieri
http://www.astro.unipd.it/quantumastronomy/documents/Fornasier05_lunarKBOocc.pdf (2005).
[21] Braga Ribas, F., Ph D Thesis, Observatorio Nacional RJ (2013).
[22] Braga Ribas, F., et al., Nature [in press, doi: 10.1038/nature13155 ] (2014)
[23] Waagen, E. O., AAVSO Circular #499 (2014)
[24] Richichi, A. and A. Glindemann, Astron. & Astrophysics, 538 A56 (2012).
[25] McAlister, H.A.; et al. , Ap. J. 628, 439–452 (2005).
[26] F. Cain, http://www.universetoday.com/10213/egg-shaped-regulus-is-spinning-fast/ (2005).
[27] Heyrovský, D., Astrophys. J., 656:483-492, (2007).
[28] Sigismondi, C., T. Flatrès, T. George, F. Braga-Ribas, Atel # 5987 (2014).
[29] Petsov, A.A., et al., Astron. Nachr. **335**, 21 – 26 (2014).
[30] George, T., http://www.asteroidoccultation.com/observations/NA/2013Meeting/R-OTE%202013%20IOTA%20Conference.pdf (2013) .
[31] Rubin, V., Astrophys. J. 129, 812 (1959).
[32] Wilson, R. E., Mon. Not. R. astron. Soc. 145, 367 (1969).
[33] Russell, H. N. and H. Shapley, Astrophys. J. 36, 239 (1912).
[34] Russell, H. N. and H. Shapley, Astrophys. J. 36, 365 (1912).
[35] Huffer, C. M., Proc. Am. Astron. Soc. 8, 244 (1936).
[36] Barnes, J. W., Astrophys. J. 705, 683 (2009).
[37] Flatrès, T., Fren_difl https://docs.google.com/file/d/0BzHjNDloaOsKTi1hejJYOXN5MlU/edit
[38] http://dstats.net/download/http://www.asteroidoccultation.com/observations/R-OTE%20Release%20Package.zip